\documentclass[twocolumn]{aa}

\usepackage[varg]{txfonts}
\usepackage{graphicx}
\usepackage[usenames,dvipsnames]{color}
\usepackage{amssymb}
\usepackage{amsfonts}
\usepackage{amsbsy}
\usepackage{amsmath}

\def\SNIa{SN\,Ia}
\def\SNeIa{SN\,Ia}

\def\wa{\ensuremath{w_{\rm a}}}
\def\sigwa{\ensuremath{\sigma_{w_{\rm a}}}}

\def\minnum{{\em HST/3yrs}}
\def\maxnum{{\em HST/6yrs}}

\begin{document}

\title{Improving Dark Energy Constraints with High Redshift Type Ia Supernovae from CANDELS and CLASH}
\titlerunning{Dark Energy Constraints with High Redshift CANDELS and CLASH Supernovae}
\authorrunning{Salzano et al.}

\author{Vincenzo~Salzano\inst{\ref{inst1}}, Steven~A.~Rodney\inst{\ref{inst2}}, Irene~Sendra\inst{\ref{inst1}}, Ruth~Lazkoz\inst{\ref{inst1}}, Adam~G.~Riess\inst{\ref{inst2},\ref{inst3}}, Marc~Postman\inst{\ref{inst3}}, Tom~Broadhurst\inst{\ref{inst1}} \and Dan~Coe\inst{\ref{inst3}}}
\institute{Fisika Teorikoaren eta Zientziaren Historia Saila, Zientzia eta Teknologia Fakultatea, Euskal
Herriko Unibertsitatea UPV/EHU,\\ 644 Posta Kutxatila, 48080 Bilbao, Spain\label{inst1}
\and
Department of Physics and Astronomy, The Johns Hopkins University, Baltimore, MD 21218, USA\label{inst2}
\and
Space Telescope Science Institute, Baltimore, MD 21218\label{inst3}}

\date{}

\abstract{}{We investigate the degree of improvement in dark energy constraints that can be achieved by extending Type Ia Supernova (\SNeIa) samples to redshifts $z>1.5$ with the Hubble Space Telescope (HST), particularly in the ongoing CANDELS and CLASH multi-cycle treasury programs.}{Using the popular CPL parametrization of the dark energy, $w=w_0+w_a(1-a)$, we generate mock \SNIa\ samples that can be projected out to higher redshifts. The synthetic datasets thus generated are fitted to the CPL model, and we evaluate the improvement that a high-z sample can add in terms of ameliorating the statistical and systematic uncertainties on cosmological parameters.}{In an optimistic but still very achievable scenario, we find that extending the HST sample beyond CANDELS+CLASH to reach a total of 28 \SNeIa\ at $z>1.0$ could improve the uncertainty in the \wa\ parameter, \sigwa, by up to 21\%. The corresponding improvement in the figure of merit (FoM) would be as high as 28\%. Finally, we consider the use of high-redshift \SNIa\ samples to detect non-cosmological evolution in \SNIa\ luminosities with redshift, finding that such tests could be undertaken by future space-based infrared surveys using the James Webb Space Telescope (JWST).}{}


\maketitle

\section{Introduction}
\label{sec:intro}

The seminal works \citep{SNfirst1,SNfirst2}
presented the first solid evidence for the present acceleration of the
universe, by extending the Hubble diagram of Type Ia Supernovae
(\SNeIa) to redshifts $z>0.5$.  The use of \SNeIa\ as distance
indicators has now become a key component of modern investigations.
However, we still lack a clear picture about the nature and the
fundamental properties of the so-called ``dark energy'' driving this
acceleration.  As a consequence, many observational and theoretical
problems posed in the last 15 years have not yet been addressed
satisfactorily.  The \textit{consensus} cosmological model,
$\Lambda$CDM \citep{LCDM1,LCDM2}, based on the well known cosmological
constant, provides a good fit to most of the data
\citep{LCDMtest1,LCDMtest2,LCDMtest3,CMB}. This model, however, is also affected by serious
theoretical shortcomings \citep{LCDMproblems}.  Many possible
alternatives have been introduced by our imaginative community,
including modifications of the theoretical framework that deviate from
General Relativity such as Chaplygin gas, Cardassian expansion, DGP brane-world
models, $f(R)$ theories, etc. \citep{altmod1,altmod2,altmod3,altmod4,altmod5,altmod6,altmod7,altmod8,altmod9}.
Amongst observers, the most popular class of alternative
models are the proposals for evolving dark energy.  There is no strong
theoretical reason for dark energy density to remain constant as the universe
expands, and simple models for dark energy evolution can already be
investigated with currently accessible cosmological probes \citep{obsrev}.

Any fundamental component of the universe can be described, among other things, by its equation of state (EoS), defined as the ratio of its pressure $p_{X}$ to its density, $\rho_{X}$: $w(a) \doteq p_{X}(a)/ \rho_{X}(a)$, where $a\doteq1/(1+z)$ is the scale factor of the universe (see for instance \citep{Weinberg}). For matter, $w=0$, for radiation, $w=1/3$, and for dark energy this EoS parameter is fundamentally unknown.  If one assumes a constant EoS, current observations place limits of $w=-1.06\pm0.07$, consistent with the cosmological constant, $\Lambda$ \citep{SNLS31,SNLS32,SNLS33}.  However, when the EoS is allowed to evolve with the expansion, observational constraints are much weaker.

A significant challenge in this effort is to devise a physically adequate expression for the evolving dark energy EoS: for it to be useful, $w(a)$ must be sufficiently sophisticated to be able to accommodate the data, and simple enough so as to provide reliable predictions. 
As the vast literature on the topic has proved repeatedly, the joint use of \SNeIa\ and other cosmological tools such as Baryon Acoustic Oscillations (BAO) \citep{BAO1,BAO2,BAO3} and the Cosmic Microwave Background (CMB) \citep{CMB} can help us clarify this
picture in the context of a particular cosmological model.

Unfortunately, the signal-to-noise ratio at present is not so significant as to provide satisfactory constraints in more than two dark energy parameters \citep{2de1,2de2}, so for that reason, conclusions do not vary significantly when one considers different parameterizations that are
close to each other in a broad sense (smoothness, slow evolution, high redshift boundedness). Not surprisingly, a representative of this class of parameterizations, the so-called Chevalier-Polarski-Linder (CPL) model \citep{CPL1,CPL2}, has become very popular. Here it will be adopted as our reference scenario. This parametrization defines the dark energy EoS as:
\begin{equation}
w(a) = w_{0} + (1-a) w_{a} \; ,
\end{equation}
where $w_{0}$ is the value of EoS at the present time, and $w_{0}+w_{a}$ is its value at $a = 0$, i.e. at redshift $z \rightarrow \infty$.  Various attempts to generalize this expression have been made, pointing out the \textit{intrinsic} difficulty to constrain the
evolutionary parameter $w_{a}$ with datasets that are limited in their redshift range (such as \SNeIa); or the strong
correlation among the two parameters $w_{0}$ and $w_{a}$ \citep{demodels1}. For a small selection of alternative parameterizations the interested reader is referred to \citep{demodels21,demodels22,demodels23,demodels24,demodels25,demodels26,demodels27,demodels28,demodels29,demodels210,demodels211}.

The problem of placing observational constraints on $w_{a}$ is perhaps
the weakest feature of this parametrization, and the findings in this
paper are particularly relevant to this point. Specifically, our
results show that a statistically significant number of \SNeIa\ at
high redshift ($z>1.0$) can provide a valuable reduction in the errors
on $w_{a}$, i.e. on the asymptotic value of the dark energy EOS.  It
is well known \citep{Riess06,CLASH} that at $z<1$ \SNeIa\ distance measurements are most
sensitive to the ``static component'' of the dark energy, $w_{0}$;
whereas at $1<z<1.5$, the behaviour gets reversed and measurements are
most sensitive to the ``dynamic component'', $w_{a}$.  At even higher
redshift, $z>1.5$, \SNIa\ measurements could be most sensitive to
peculiarly divergent evolution in the EoS or systematic changes in the \SNeIa\ themselves (if present)
\citep{Riess06}. Thus, extending
\SNeIa\ observations to a higher redshift range than currently
available (the maximum redshift in current samples is $z_{max} \sim 1.39$) could be a valuable
step toward improving dark energy constraints.

In 2010, the Hubble Space Telescope embarked on three ambitious
Multi-Cycle Treasury (MCT) programs, designed to span three years and
produce a lasting archive of deep multi-wavelength imaging. Two of
these programs take advantage of the infrared survey capabilities of the new Wide Field Camera 3 (WFC3) to enable the
discovery and follow-up of \SNeIa\ out to $z\sim2.3$: the Cosmic
Assembly Near-infrared Deep Extragalactic Legacy Survey (CANDELS, PIs:
Faber and Ferguson) \citep{Grogin:2011a}, and the Cluster Lensing and
Supernova survey with Hubble (CLASH, PI: Postman) \citep{CLASH}.  The
CANDELS+CLASH SN search program (PI: Riess) comprises the SN
search component from both programs.  This SN survey is principally
aimed at high redshift SNe ($z>1.5$) in order to measure the time
dependence of the dark energy equation of state and improve our
understanding of \SNIa\ progenitor systems.

Our goal in this work is to measure the degree of influence that these high-z \SNeIa\ could exert in constraining dark energy parameters.  We adopt the CPL parametrization as a good representative of modern dark energy models \citep{DETF,JDEM}, and we use the \textit{SuperNova Legacy Survey} compilation (SNLS3) \citep{SNLS31,SNLS32,SNLS33} as the basis for comparison against future high-z samples.

In Section II we describe the method we followed to simulate the expected final set from CANDELS+CLASH, and a larger sample representing an extended high-z \SNeIa\ survey with HST. Finally, in Section III we discuss the results expected for future dark energy constraints and explore some other interesting features that can be achieved only with high-z \SNeIa.

\section{Mock Data Algorithm}

To study the impact of high redshift \SNeIa\ on dark energy constraints, we want to extend the SNLS3 samples by adding a synthetic sample of high-z \SNeIa\ that mimics what can be done with HST. In this, we proceed using, in parallel, mock simulations of high redshift \SNeIa\ and the Fisher matrix formalism. We first realize a Mark Chain Monte Carlo (MCMC) analysis, assuming a diagonal covariance matrix and a fixed $\Omega_{m}$ with our mock data sets (the reasons behind this choice are detailed in the following sections). Then, we perform a Fisher matrix analysis on the same data sets and with the same conditions, thus obtaining a proportionality factor to convert Fisher-derived errors into MCMC-derived ones. As it is well known, the Fisher matrix formalism \citep{Fisher1,demodels21} only provides a minimum error forecast on the considered quantities as it is based on the ideal hypothesis of Gaussian errors. On the other hand, creating mock data and analyzing them gives more realistic estimations for cosmological constraints \citep{Fisher2}. We have thus compared errors on the interested parameters derived from a statistical analysis of mock data with those ones estimated through the Fisher formalism; we have been able to find a reliable proportionality between them. This is a quite useful result, as the Fisher matrix is a more efficient procedure than the MCMC method, and can be easily implemented for any number of \SNeIa\ one wants to consider. Finally, we improve on those initial results with an MCMC analysis performed on the real SNLS3 data set, where we now utilize the full covariance matrix, and allow $\Omega_{m}$ to be a free parameter, constrained by appropriate priors. Comparing this more rigorous analysis to the results from the previous steps , we find a scaling factor to correct the (underestimated) cosmological parameter errors into more realistic estimates.

To begin, we first generate our mock \SNeIa\ data, using the following procedure:
\begin{enumerate}
 \item Fit a CPL cosmological model to the real SNLS3 dataset; this one will be used as the fiducial cosmological background to produce the mock samples;
 \item Simulate all the observational quantities and their related errors that enter the SNLS3 data set (for more details, see below in this section), giving us a mock SNLS3 sample;
 \item Check the accuracy of the simulation procedure by comparing this mock sample with the real one, also performing a cosmological fit with it;
 \item Use the same algorithm to create the high-z HST mock samples we need, mimicking the expected CANDELS+CLASH yield;
\end{enumerate}

One should expect estimates for the EoS parameters to improve in precision as the sample grows. As discussed in the following section, this
procedure for generating mock \SNeIa\ data is designed to avoid systematic biases. Thus, any improvement in precision can be completely attributed to the high redshift \SNeIa.

Finally, the information used to generate the mock SNLS3 data set (step 2) can be utilized in the Fisher matrix formalism. For a given \SNIa\ sample, the Fisher matrix approach gives a prediction for the errors on cosmological parameters of interest.  We then compare these (underestimated) Fisher matrix errors to the more realistic errors derived from the mock data analysis.  As described in Section 2.4, this comparison reveals a tight proportionality that can be used to improve the Fisher matrix results.

\subsection{SNLS3 preliminary considerations}

We first turn our attention to step 1: fitting the SNLS3 data to define our fiducial CPL model. The quantity of interest in the SNLS3 context is the predicted magnitude of the \SNeIa\ $m_{\rm mod}$, which describes the relative light-curve for any \SNeIa\, given the cosmological model and two other quantities,  the stretch (a measure of the shape of the \SNeIa\ light-curve) and the color. It reads
\begin{equation}\label{eq:m_snls3}
m_{\rm mod} = 5 \log_{10} [ d_{L}(z, \Omega_m; \boldsymbol{\theta}) ] - \alpha (s-1) + \beta \mathcal{C} + \mathcal{M} \; .
\end{equation}
where $\alpha$ and $\beta$ are the parameters which characterize the stretch-luminosity and color-luminosity relationships, reflecting the well-known broader-brighter and bluer-brighter relationships, respectively \citep{SNLS32}.

It depends on $d_{L}(z, \Omega_m; \boldsymbol{\theta})$, the Hubble free luminosity distance:
\begin{equation}\label{eq:dl_H}
d_{L}(z, \Omega_m; \boldsymbol{\theta}) = (1+z) \ \int_{0}^{z}
\frac{\mathrm{d}z'}{E(z', \Omega_m; \boldsymbol{\theta})} \; ,
\end{equation}
with the Hubble function $E(z) = H(z)/H_{0}$ depending on the matter and dark energy components and $\boldsymbol{\theta}$ being the EoS parameters vector ($\boldsymbol{\theta} = (w_{0}, w_{a})$ for the CPL cosmological model).  $\mathcal{M}$ is a nuisance parameter combining the Hubble constant $H_{0}$ and the absolute magnitude of a fiducial \SNeIa. This notation is chosen to clarify the roles of the various cosmological parameters appearing in the formulae. The fractional matter density parameter $\Omega_m$ is separated as we fix it to an a priori value ($\Omega_{m} = 0.259$, derived from a quiessence model analysis of WMAP7.1 CMB data\footnote{http://lambda.gsfc.nasa.gov/product/map/current/params/\\wcdm$\_$sz$\_$lens$\_$wmap7.cfm}).

By assuming a fixed $\Omega_{m}$ rather than leaving it as a free parameter we can vastly simplify the generation of our mock data sets.  This is
important because, as we will discuss in the next section, we have a large number of parameters contributing to the total error budget that we have to simulate.  By fixing $\Omega_{m}$ we can focus our analysis on observing the effects in $w_{0}$ and $w_{a}$.  Unfortunately, this assumption necessarily leads to an underestimation of the errors for these dark energy parameters.  In Section 2.4 we describe the procedure that we have implemented to address this shortcoming, which recovers a more realistic estimation of the $w_{0}$ and $w_{a}$ uncertainties.

Finally, $\boldsymbol{\theta}$ is the EoS parameters vector ($\boldsymbol{\theta}=(w_{0}, w_{a})$ for the CPL one), which we will have to constrain.

Eqs.~(\ref{eq:m_snls3})~-~(\ref{eq:dl_H}) can be used after defining a cosmological model to be tested. As we will fit our \SNeIa\ data with the CPL model, the Hubble function $E(z) = H(z)/H_{0}$ will depend on matter and dark energy components through the expressions
\begin{equation}
E(z) = \left[ \Omega_{m} (1+z)^{3}+ (1-\Omega_{m})X(z,\boldsymbol{\theta}) \right]^{1/2} \; ,
\end{equation}
and
\begin{equation}
X(z,\boldsymbol{\theta})= \exp \left[3\int_{0}^{z} \frac{dz'}{1+z'} (1+w(z',\boldsymbol{\theta})) \right] \; .
\end{equation}
Here we have assumed spatial flatness. Using the CPL parametrization we get:
\begin{equation}
X(z)= (1+z)^{3(1+w_{0}+w_{a})} \exp \left[-\frac{3 w_{a} z}{1+z}\right].
\end{equation}

There is an important difference in the SNLS3 handling of the parameters $\alpha$ and $\beta$ which enter in Eq.~(\ref{eq:m_snls3}) with respect to other \SNeIa\ samples in the literature, such as the Union2 data set \citep{Union2}. In the SNLS3 approach these parameters are left free during all stages of their error estimation, both for statistical and systematic uncertainties; in other cases \citep{Union2}, the authors find best-fit values for $\alpha$ and $\beta$ in the preliminary stages, when building the \SNeIa\ sample, and then hold $\alpha$ and $\beta$ at those fixed values for uncertainty calculations in the cosmological analysis. In \cite{SNLS32} the authors demonstrate that this latter approach can introduce a bias in the determination of the cosmological parameters.

Furthermore, the SNLS3 team\footnote{http://www.cfht.hawaii.edu/SNLS/} advocates the use of all the components (both statistical and systematic errors) of the covariance matrix.  Most importantly for our purposes, they provide data files with the full multidimensional covariance matrix for all the physical quantities involved in their analysis, allowing us to reconstruct the contributions of first-order covariance terms to the uncertainties (see Section 2.1). As the main purpose of this work is to detect the relative decrease of errors (if there is any) on cosmological parameters by adding simulated high redshift \SNeIa\ from HST, we made two choices: \textit{(1.)} we have left $\alpha$ and $\beta$ free as prescribed in \cite{SNLS32}; \textit{(2.)} we use only diagonal components of the statistical and systematic covariance matrix, because of the intrinsic difficulty in simulating out-of-diagonal terms of the covariance matrix (see Section 2.2). In this way, the expected underestimation of real errors on cosmological parameters \citep{SNLS33} due to the sole use of statistical observational errors is highly softened. Moreover, this does not affect the primary goal of this work, which is to quantify the \textit{relative} decrease of errors on cosmologically interesting dark energy parameters.

\subsection{Generating mock data and errors}

To generate our mock SNLS3 catalog of $m_{\rm mod}(z_{i})$ values, we use:
\begin{equation}
{m_{\rm mod}}(z_{i}) = {m_{\rm mod, CPL}}(z_{i}) + \Delta {m_{\rm mod, CPL}}(z_{i}) \, ,
\end{equation}
where: $z_{i}$ with $i=1, \ldots, \mathcal{N}_{data}$ and $\mathcal{N}_{data} = 472$ for SNLS3, are randomly drawn redshift values selected to match the true redshift distribution in the real \SNeIa\ samples; ${m_{\rm mod, CPL}}$ is the value of the magnitude at a given redshift calculated from the best fit of SNLS3 sample with a CPL model. It is important to underline here that in order to calculate the fiducial ${m_{\rm mod, CPL}}$ value we have to simulate also the stretch and color quantities. The $\alpha$ and $\beta$ parameters are derived from the best fit analysis; and the nuisance parameter $\mathcal{M}$ can be calculated following \cite{SNLS32}.

The offset $\Delta{m_{\rm mod, CPL}}$ is a random noise component added to ${m_{\rm mod, CPL}}$ in order to give the mock dataset a realistic dispersion (i.e. similar to the real datasets). This random noise is sampled, for any redshift, from a Gaussian distribution with zero mean and standard deviation $\sigma_{m_{B}}^{eff}$ (this is defined below in this section as the total magnitude-only error).

\begin{figure*}
\centering
\includegraphics[width=17.5cm]{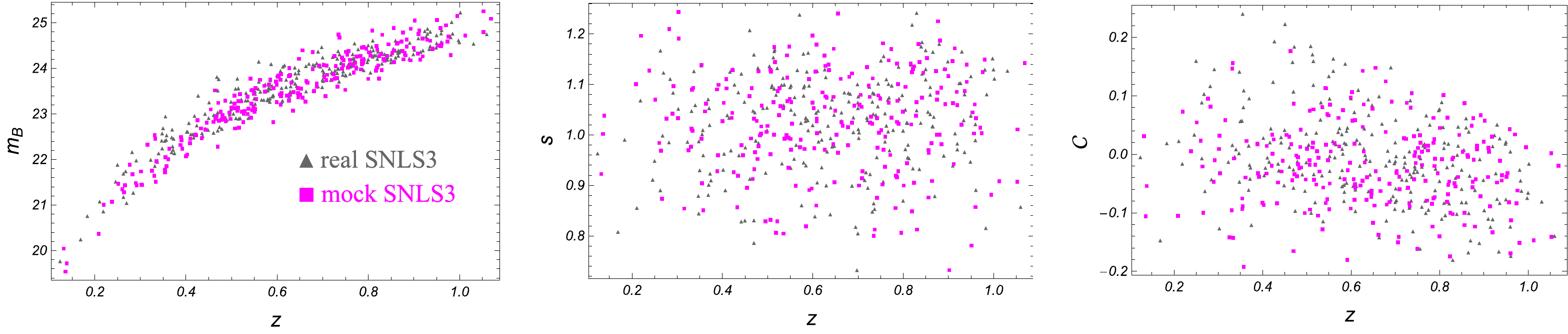}\\
~~~~~\\
\includegraphics[width=18.cm]{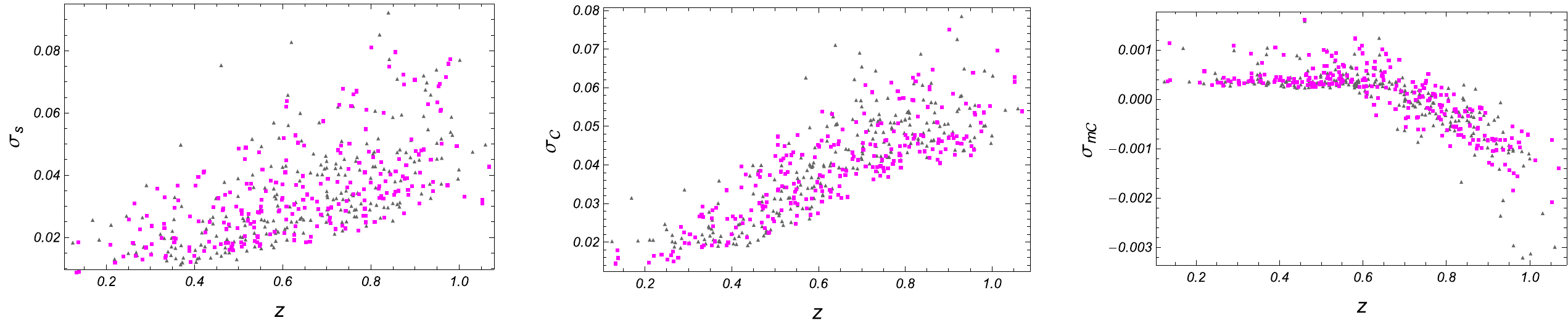}\\
\caption{Comparing real and mock SNLS sub-sample from the SNLS3 data set: light-gray triangles are for real SNLS3 data, magenta squares for mock SNLS3 data. \textit{Top.} From left to right: magnitude, stretch, color vs redshift. \textit{Bottom.} From left to right: diagonal stretch error, color error and magnitude-color covariance vs redshift.} 
\label{fig:mock_Union_err1}
\end{figure*}

\begin{figure*}
\centering
\includegraphics[width=7.8cm]{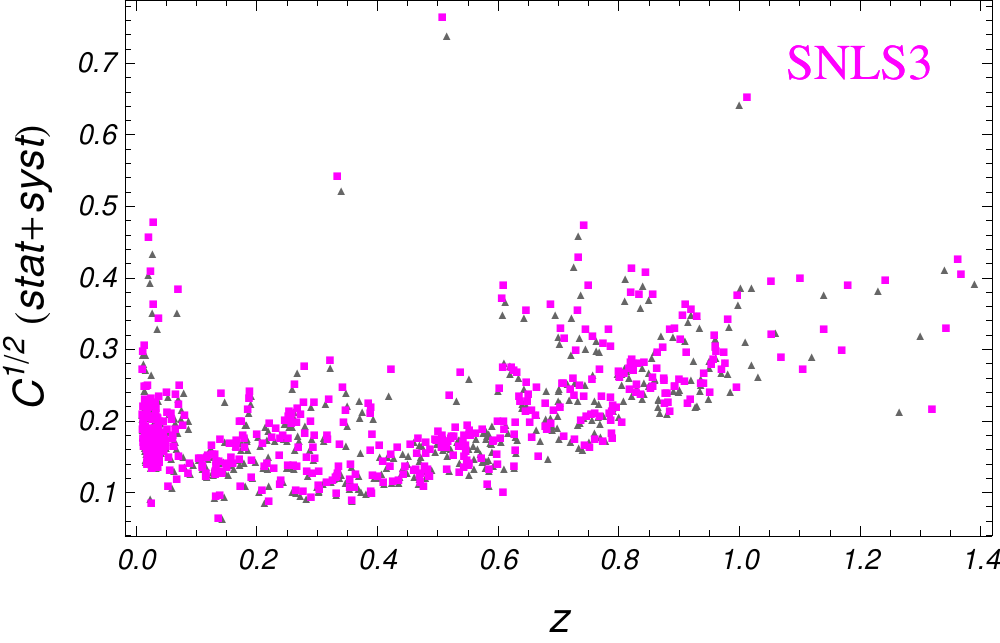}~~~~~
\includegraphics[width=7.8cm]{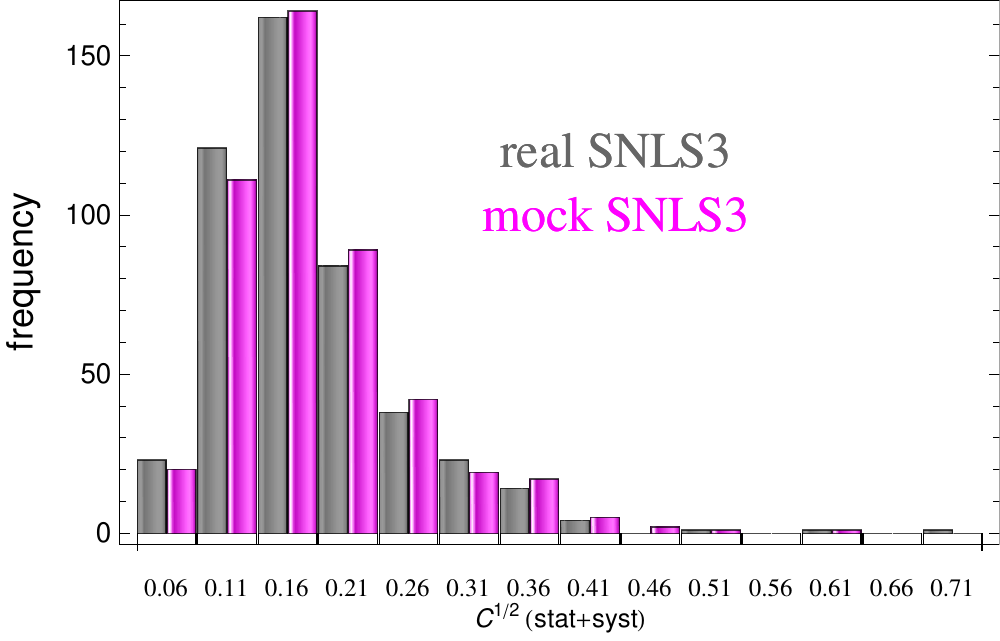}\\
\caption{\textit{Left.} Plot of total statistical plus systematic diagonal errors on \SNeIa\ magnitude vs redshift: light-gray triangles are for real SNLS3 data, magenta squares for mock SNLS3 data. \textit{Right.} Comparison of histograms for the same errors from the real and mock SNLS3 datasets (gray and magenta bars, respectively).}
\label{fig:mock_Union_err2}
\end{figure*}

While it is relatively easy to produce mock values for the \SNIa\ magnitudes, more attention must be paid to the reproduction of errors on
$m_{\rm mod}(z_{i})$. Here we emphasize again that we are using only the diagonal part of the total covariance matrix. The statistical errors on the magnitude $m_{\rm mod}$ are given by the relation \citep{SNLS32}:
\begin{eqnarray}
\sigma_{stat}^2 &=& \sigma_{m_{B}}^{eff,2} + \alpha^2 \sigma_{s}^2 + \beta^2 \sigma_{\mathcal{C}}^2 + 2 \alpha \sigma_{ms} - 2 \beta \sigma_{m\mathcal{C}} + \nonumber \\
&-& 2 \alpha \beta \sigma_{s\mathcal{C}} + \sigma_{z}^2 \; ,
\label{eq:snlserror}
\end{eqnarray}
with
\begin{equation}
\sigma_{m_{B}}^{eff,2} = \sigma_{int}^2 + \sigma_{pec}^2 + \sigma_{m_{B}}^{2} \; .
\end{equation}
The different quantities contributing to $\sigma_{stat}$ are described as follows: $\sigma_{m_B}$ is the error on the observed magnitude including lensing and host galaxy effects; $\sigma_{s}$ and $\sigma_{\mathcal{C}}$ are the errors on the stretch and color parameters; $\sigma_{ms}$, $\sigma_{m\mathcal{C}}$ and $\sigma_{s\mathcal{C}}$ are the covariance terms between magnitude, stretch and color; $\sigma_{int}$ is the intrinsic scatter of \SNeIa\ and it is different for any sub-sample; $\sigma_{pec}$ is the peculiar velocity error; $\sigma_{z}$ is the error on redshift converted into magnitude uncertainty. See \citep{SNLS32} for more details on their calculation.

This is only a part of the total covariance matrix; taking into account also the systematic terms, this matrix would be:
\begin{eqnarray}
\mathbf{C} &=& \boldsymbol{\sigma_{stat}^2} + \boldsymbol{\sigma_{syst,m_{B}}^2} + \alpha^2 \boldsymbol{\sigma_{syst,s}^2} + \beta^2 \boldsymbol{\sigma_{syst,C}^2} + \nonumber \\
&+& 2 \alpha \boldsymbol{\sigma_{syst,ms}^2} - 2 \beta \boldsymbol{\sigma_{syst,mC}^2} - 2 \alpha \beta \boldsymbol{\sigma_{syst,sC}^2} \; ,
\label{eq:systerror}
\end{eqnarray}
where the bold font is used for matrices, the suffixes refer to the statistical or systematic nature of the errors, and other symbols have the same meaning as in Eq.~(\ref{eq:snlserror}).

The difficulty to build mock error samples lies in the dependence of the statistical errors on the free parameters $\alpha$ and $\beta$; for this reason we cannot attempt a global phenomenological fit of them. To fully describe the total statistical and systematic errors we have to generate mock sub-samples survey by survey (as they are given in Table~3 of \cite{SNLS32}) for all the terms on the right sides of Eqs.~(\ref{eq:snlserror})~-~(\ref{eq:systerror}). For most of the SNLS3 sub-samples, we find a good fit for the error terms by assuming a power law $\propto {(1+z)}^n$, generally obtaining $n\sim1$ or $\sim2$. However, in those cases where no clear pattern was detectable, we have opted for a random distribution (independent of redshift) which closely mimics the real data. For the $\sigma_{m_{B}}^{2}$ term, which contains some redshift dependent elements, we have performed fits following the relation given by \citep{snerror4}: $\sigma_{s}^2 \left(\frac{1+z}{1+z_{max}}\right)^2$, with $z_{max}$ being the maximum observable redshift for each survey.

\subsection{Testing the goodness of mock-building algorithm}

Despite the high number of parameters entering the definition of the total covariance matrix, we can see in Figs.~\ref{fig:mock_Union_err1}~-~\ref{fig:mock_Union_err2} that our simple empirical approach does yield mock quantities and errors that match the true distributions quite reliably (only for illustrative purposes, in Fig.~\ref{fig:mock_Union_err2} we have fixed $\alpha$ and $\beta$ to the best fit values from the cosmological fit of real SNLS3).

\begin{figure}[htbp]
\centering
\includegraphics[width=7.7cm]{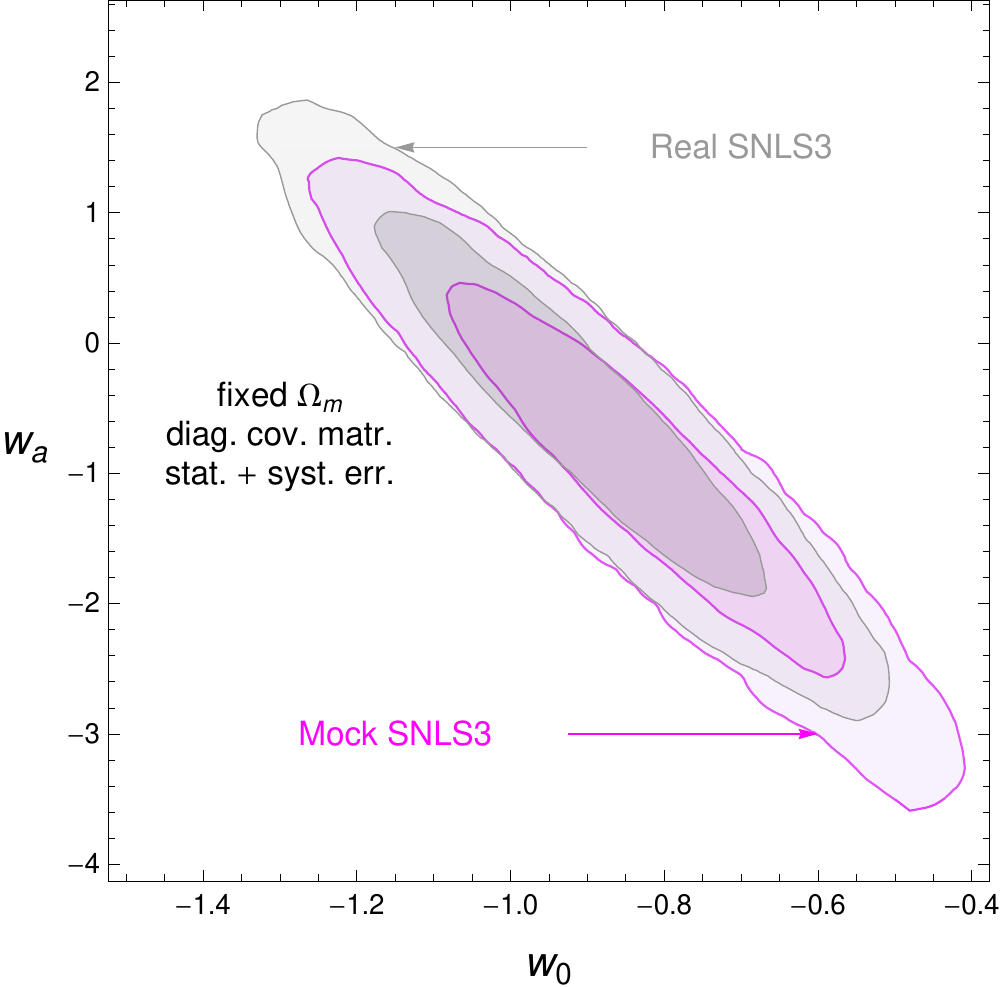}
\caption{Fitting the CPL model with \SNIa\ data: $1\sigma$ and $2\sigma$ contours in the $w_{0}-w_{a}$ parameter space. Gray contours are from
  real SNLS3 data set; magenta contours are from mock SNLS3 data set. Contours reflect statistical plus systematic errors.} \label{fig:Union_mock_vs_real}
\end{figure}

In order to firmly validate the quality of our mock procedure, we also compare the outcomes of cosmological fits to real and mock SNLS3 data sets. These fits have been obtained by minimizing the quantity:
\begin{equation}\label{eq: sn_chi_snls3}
\chi^{2}(\boldsymbol{\theta}) = \sum^{\mathcal{N}_{\mathrm{SN}}}_{j =
1} \frac{(m_{B}(z_{j}) - m_{\rm mod}(z_{j}, \Omega_m; \alpha, \beta, \boldsymbol{\theta}))^{2}}{C_{j}} \; .
\end{equation}
We marginalize over the nuisance parameter $\mathcal{M}$ following the prescription given in \cite{SNLS32}. Recent empirical evidence suggests two different values of $\mathcal{M}$ depending on the value of the host stellar mass: so we define two sub-samples, one with $\log M_{host} < 10$ and the other with $\log M_{host} > 10$. The expression for this $\chi^2$ marginalized over $\mathcal{M}$ is given in Appendix C of \cite{SNLS32}.  To minimize the $\chi^2$ we use the Markov Chain Monte Carlo Method (MCMC) and test for convergence with the method described by \cite{Dunkley05}.

The comparison between the real and mock SNLS3 in Fig.~\ref{fig:Union_mock_vs_real} shows a slight shift in both cosmological parameters, although the errors are quite the same. This is the price to be paid (in addition to the usual problems of mock procedures) for not fitting the total errors directly: as we have discussed in the previous section, we have to fit each term contributing to the error and each of them contributes with its dispersion to the final results. Moreover, we have two more parameters, $\alpha$ and $\beta$ in the fit, which can influence the final results by interacting with the stretch and the color parameters and with the error expression. But we underline that an \textit{absolute} comparison of results is beyond the scope of this work; what is important is to verify that the cosmological parameter uncertainties are of the same order (if not almost perfectly coincident) and the contours are quite equivalent regardless of the small shift. A way to quantify this strong agreement is the FoM, defined as the inverse of the area of the contours:  for the mock SNLS3 data set we get a FoM $\lesssim 4\%$ smaller than its real counterpart (fourth column in Table.~\ref{tab:bestfit3}). This result validates our approach and is important for three main reasons: first, it means that our mock observational quantities and errors are not losing any information that may be included in the real data; second, it justifies the generation of the synthetic CANDELS+CLASH sub-sample of \SNeIa\ by the same procedure; third, our data can be appropriately used as input in the Fisher formalism we will discuss in the next section.

\subsection{CLASH+CANDELS mock}

{\renewcommand{\tabcolsep}{1.75mm}
{\renewcommand{\arraystretch}{1.5}
\begin{table}[htbp]
\begin{minipage}{0.475\textwidth}
\caption{Mock samples.}\label{tab:clash_num}
\centering
\resizebox*{0.95\textwidth}{!}{
\begin{tabular}{ccc}
\hline \hline
Redshift range & \multicolumn{2}{c}{Number of \SNIa \footnote{Number of cosmologically useful \SNIa\ expected to be discovered by the 3-years CANDELS+CLASH program and an imagined extension into a 6-years program with HST+WFC3, including CANDELS+CLASH and another similar 3-year survey.}} \\
\hline
 & CANDELS+CLASH & HST+WFC3 6yr \\
 & (\minnum) & (\maxnum) \\
\hline \hline
$1.0 < z < 1.5$ & $8$ & $16$ \\
$1.5 < z < 2.0$ & $5$ & $10$ \\
$2.0 < z < 2.5$ & $1$ & $2$ \\
\hline
Total: & $14$ & $28$\\
\hline\hline
\end{tabular}}
\end{minipage}
\end{table}}}

In Table~\ref{tab:clash_num} we define two plausible distributions of high redshift \SNeIa\ events that could be observed with HST+WFC3. The ``\minnum'' mock sample is designed to mimic the expected yield from the 3-year \SNeIa\ survey of CANDELS+CLASH, with 14 \SNeIa\ at high redshift (i.e. at $z>1$) out to $z=2.5$ that are useful for cosmology (i.e. with light curves).  The ``\maxnum'' mock sample imagines some future survey with WFC3 that extends the HST \SNeIa\ detections for another 3 years, doubling the sample to 28 \SNeIa\ out to $z=2.5$.

For each of these samples we must extrapolate from existing data to predict the distance modulus uncertainties in the new redshift regime. We extrapolate from the SNLS3 sub-sample containing $17$ \SNeIa\ observed with HST (using the ACS camera). In this way, statistical and systematic errors in the mock sample should represent a realistic yet conservative scenario. This final set of two mock samples now allows us to efficiently study the impact that a high-$z$ \SNeIa\ sample from HST could have in constraining the dark energy EoS.

These \minnum\ and \maxnum\ mock data sets are the two main tools we use for our analysis. We will also consider larger samples (up to ~100 high-redshift \SNeIa) of the same kind in order to discern if there is a simple relationship between the cosmological parameter errors that come from mock \SNeIa\ samples and those derived through the Fisher matrix formalism. As is well known, the inputs for a Fisher procedure are: a fiducial cosmological model, the redshift distribution of the sample, and the error prescriptions for the survey(s) being considered.  For the fiducial cosmological model we use the best-fit CPL model from our fit to the real SNLS3 data. The redshift distribution corresponds to the redshift values from the real SNLS3, with extensions to higher redshift as given in Table~\ref{tab:clash_num}. The errors are defined by the real SNLS3 data and the mock \minnum\ and \maxnum\ samples.

With these inputs, the Fisher matrix procedure returns an estimate of the errors on the cosmological parameters of interest (the ``Fisher errors'' on $w_0$ and $w_a$). We then compare these to the errors derived from fitting a CPL model to the mock data (the ``mock errors''). We find that the ratio of Fisher errors to mock errors is quite constant, for all sample sizes (as long as the errors on cosmological parameters are very likely gaussian and with very small asymmetries).  If we define such a constant as $p_{i} \doteq \frac{\sigma^{mock}_{i}}{\sigma^{Fisher}_{i}}$ (with $i = w_{0}$, $w_{a}$) we have: $<p_{w_{0}}> = 1.59$ and $<\Delta p_{w_{0}}> = 0.03$; $<p_{w_{a}}> = 1.15$ and $<\Delta p_{w_{a}}> = 0.02$. This result allows us to scale up all the Fisher errors (which are understood to be lower limits on the true uncertainty) into a ``more realistic'' estimation of the $w_0$ and $w_a$ uncertainties, but not yet ``completely realistic''.

As noted in Sections 2.1 and 2.2, we expect that these uncertainties are still significantly underestimated because of two key simplifications we have used: assuming a fixed $\Omega_{m}$, and leaving out the off-diagonal elements of the covariance matrix.  To see how these choices influence our error estimates, we have performed an MCMC analysis of the real SNLS3 data set with the total covariance matrix given by \citep{SNLS32}. Here we have not fixed $\Omega_{m}$, but instead apply a gaussian prior on the matter content, $\Omega_{m} = 0.26 \pm 0.02$ \citep{Wang11}. This prior includes information from external cosmological data sets other than \SNeIa\ (columns 5 and 6 of first line in Table.~\ref{tab:bestfit3}).  We then repeat the analysis on the real SNLS3 data, but now using our simplification of a fixed $\Omega_{m}$ and only diagonal covariance terms. Comparing the $w_0$ and $w_a$ errors returned from these two approaches gives us an estimate of the systematic bias in $w_0$ and $w_a$ that may be introduced by our simplifying assumptions. From this comparison we derive a proportionality factor, which we then apply as a multiplicative correction for $\sigma_{w_0}$ and $\sigma_{w_a}$. In this way we are left with estimations of the dark energy EoS uncertainties that are much more realistic and reliable.

{\renewcommand{\tabcolsep}{1.75mm}
{\renewcommand{\arraystretch}{1.5}
\begin{table}
\begin{minipage}{0.5\textwidth}
\caption{\textit{SNLS3}: errors for the CPL cosmological model using the assumptions and the datasets described in the text. Results on each parameter follow from marginalization over the other parameters of the model.
  \textit{Column 1:} used dataset.
  \textit{Column 2:} $1\sigma$ confidence level for $w_{0}$.
  \textit{Column 3:} $1\sigma$ confidence level for $w_{a}$.
  \textit{Column 4:} Figure of Merit for any dataset.}
  \label{tab:bestfit3}
\centering
\resizebox*{0.65\textwidth}{!}{
\begin{tabular}{c|ccc}
\hline \hline
 & \multicolumn{3}{c}{prior $\Omega_{m}$ $+$ full cov.} \\
\hline \hline
SNLS3 & $\sigma_{w_{0}}$ & $\sigma_{w_{a}}$ & FoM \\
\hline \hline
real & $0.213$ & $1.647$ & $4.92$ \\
mock & $0.224$ & $1.722$ & $4.49$ \\
\hline
real+\minnum & $0.198$ & $1.430$ & $5.91$ \\
\hline
real+\maxnum & $0.187$ & $1.305$ & $6.82$ \\
\hline
real+$N_{z>1}=56$ & $0.177$ & $1.182$ & $7.85$ \\
\hline
real+$N_{z>1}=126$ & $0.164$ & $1.010$ & $9.51$ \\
\hline
real+\maxnum+JWST & $0.171$ & $1.123$ & $8.93$ \\
\hline \hline
\end{tabular}}
\end{minipage}
\end{table}}}

\begin{figure*}[htbp]
\centering
\includegraphics[width=7.7cm]{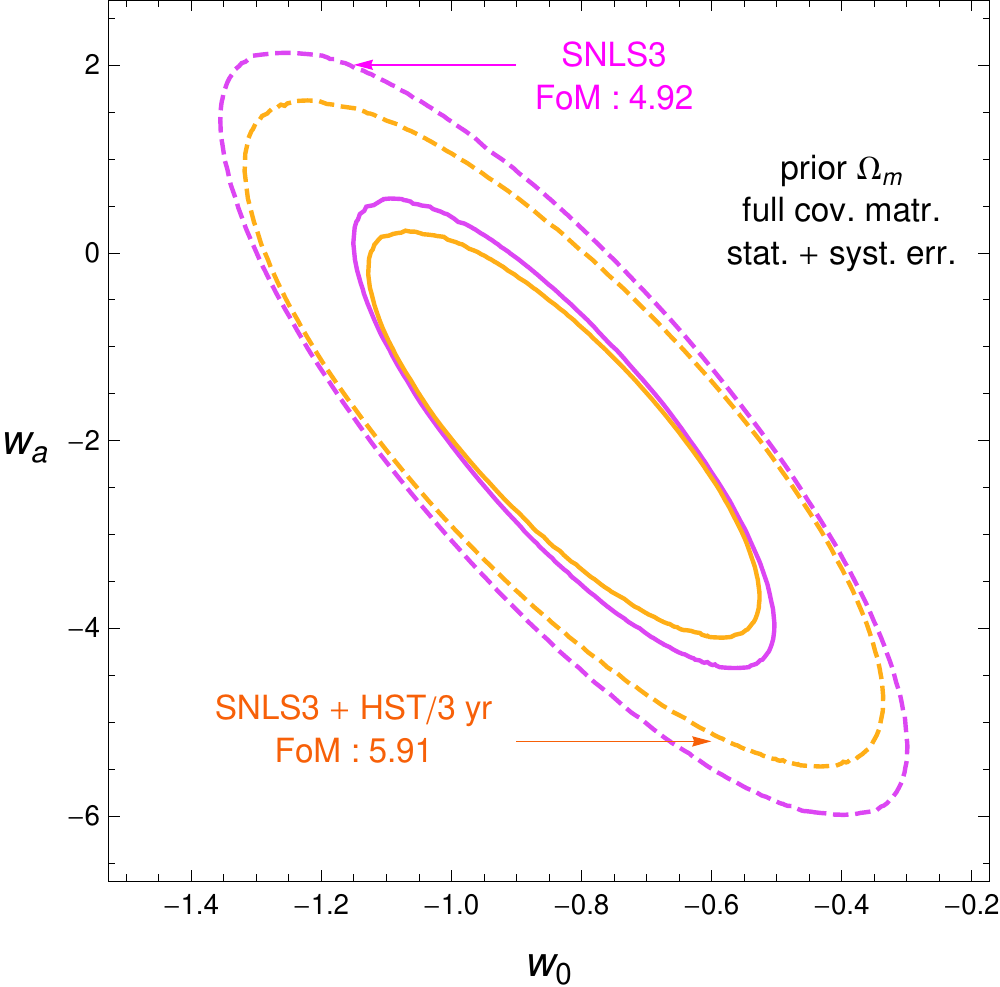}~~~~~
\includegraphics[width=7.7cm]{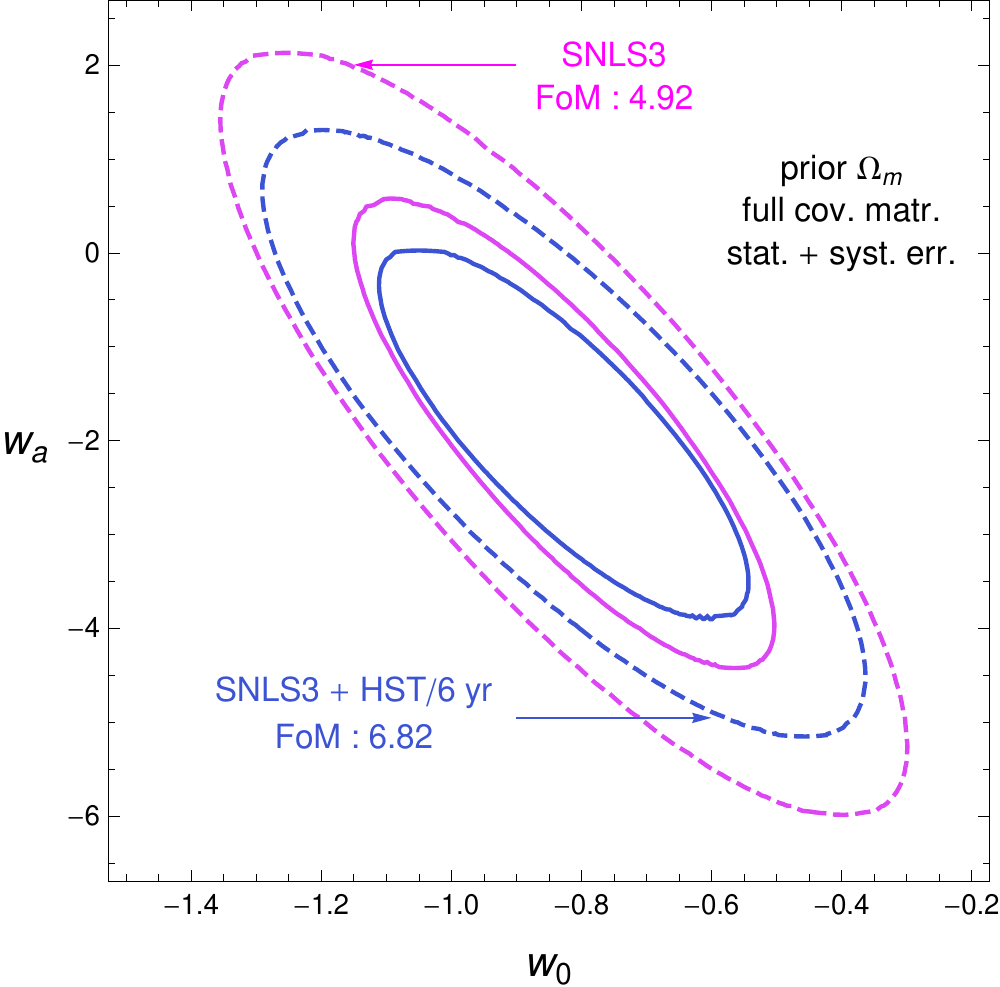}
\caption{CPL model with \SNIa\ data: $1\sigma$ and $2\sigma$ contours in the $w_{0}-w_{a}$ parameter space. \textit{(Left Panel.)} Magenta contours come from the SNLS3 data set; red contours come from the SNLS3+{\minnum} mock data set. \textit{(Right Panel.)} Magenta contours come from the SNLS3 data set; blue contours come from the SNLS3+{\maxnum} mock data set).}
\label{fig:Union_vs_Clash}
\end{figure*}

\section{Results and Conclusions}

Let us outline our main findings, keeping in mind that our fits have been carried out under the assumption of a CPL model. Fig.~\ref{fig:Union_vs_Clash} and Table~\ref{tab:bestfit3} give a quantitative summary, showing that, as expected, the high redshift \SNeIa\ make errors on the EoS parameters decrease in all cases. This effect is small for $w_{0}$, which is already quite narrowly constrained by present data: we see a reduction in $\sigma_{w_{0}}$ $\approx 7 - 12 \%$. However the improvement is more pronounced in the case of the dark energy evolution parameter $w_{a}$ with a reduction in $\sigma_{w_{a}} \approx 13 - 21 \%$. The narrower constraints on $w_{0}$ and $w_{a}$ induce in turn a significant improvement in the FoM, increasing by $20 - 39 \%$.

\subsection{Future prospects}

\begin{figure*}[htbp]
\centering
\includegraphics[width=7.7cm]{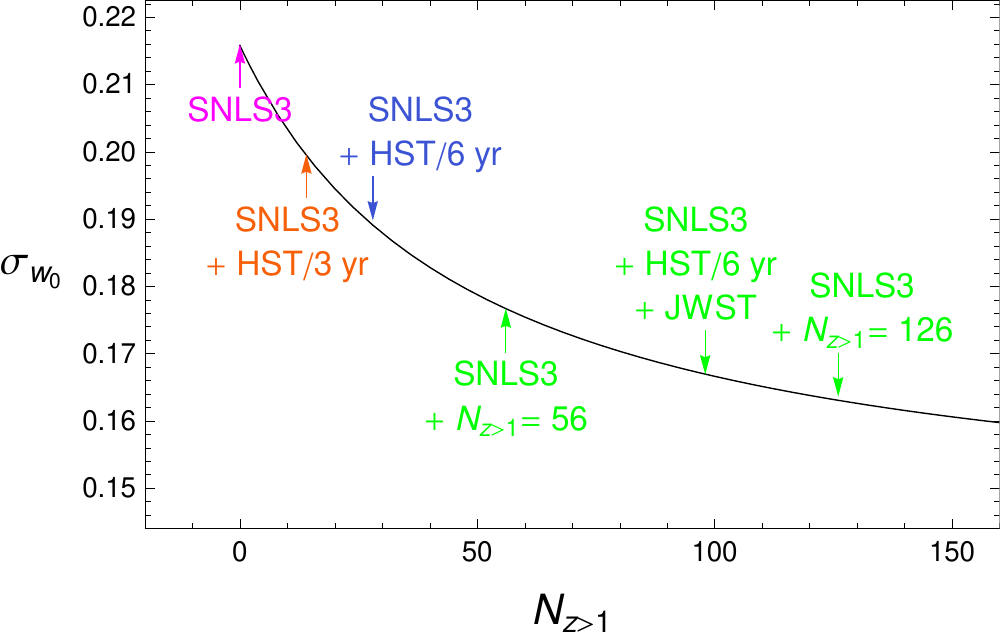}~~~
\includegraphics[width=7.7cm]{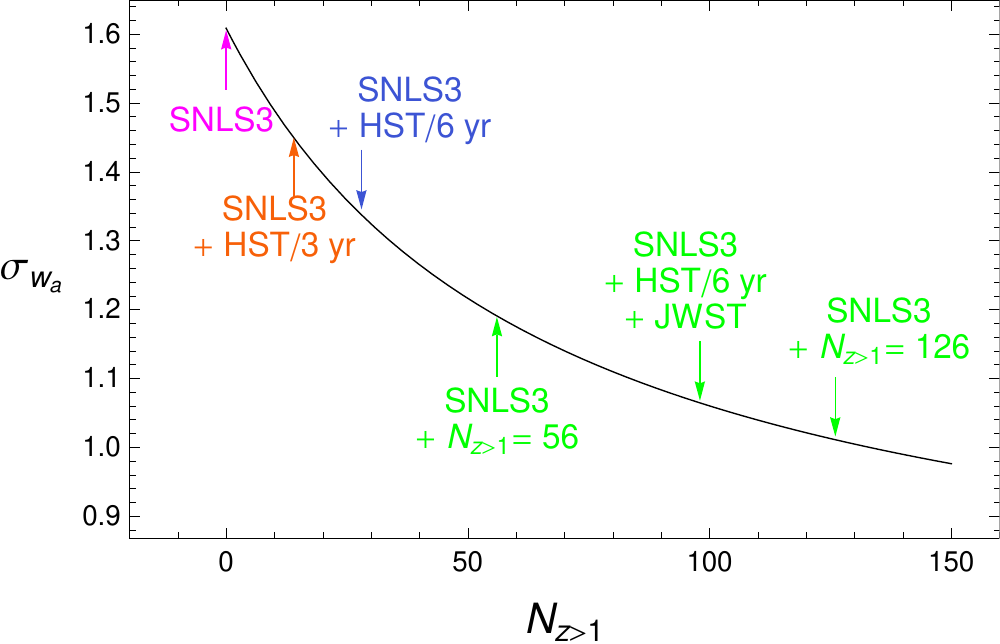}\\
~~~\\
\includegraphics[width=7.7cm]{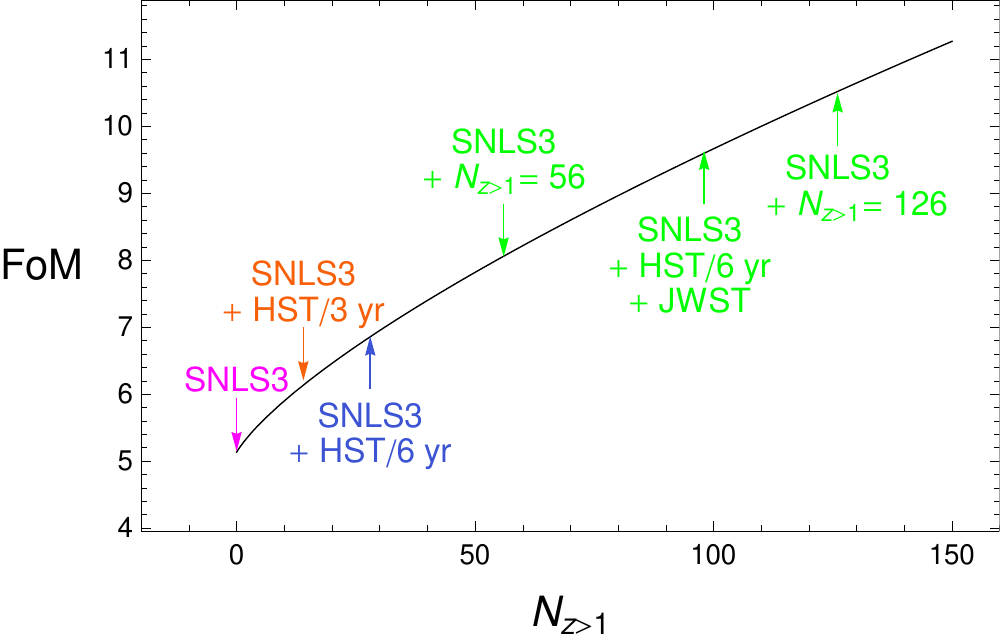}
\caption{Future perspective analysis of \SNIa\ data. $1\sigma$ errors in the $w_{0}-w_{a}$ parameter vs number of high-redshift \SNIa\ : magenta - present SNLS3 data set; red - mock SNLS3+{\minnum}-like data set; blue - mock SNLS3+{\maxnum}-like data set; green - forecasts for various number of high redshift \SNIa.}\label{fig:fisher}
\end{figure*}

We can now look beyond the immediate horizon of the current CANDELS+CLASH HST surveys, and examine how much improved precision in the measurement of $w_0$ and $w_a$ can plausibly be achieved by adding more high-z \SNeIa. We can also ask a broader question: How many high-z \SNeIa\ from space-based surveys would be needed to provide a noticeable improvement in our measurement of these cosmological parameters?

Final results (Fig.~\ref{fig:fisher}) show that adding high redshift \SNeIa, with the CANDELS+CLASH redshift distribution we have considered in this work, can produce a sensitive decrease of the statistical and systematic errors on $w_{0}$ and $w_{a}$ up to a maximum of $\mathcal{N}_{z>1} \sim 56 $. This approximately corresponds to a 10-year campaign in the mold of CLASH+CANDELS (almost certainly exceeding the remaining lifetime of HST).  This imagined sample of 56 \SNeIa\ would decrease the uncertainty in $w_0$ by $\approx 5 \%$ improvement relative to the \maxnum\ benchmark simulation (equivalently, a $\approx 17 \%$ improvement relative to the present status). For $w_a$ the projected improvement is $\approx 7 \%$ relative to the \maxnum\ expectation (or $\approx 28 \%$ relative to the present status). Beyond this point the rate of decrease of the errors starts to slow down. To obtain another reduction in errors of the same magnitude, one would need to more than double the high-z \SNeIa\ sample ($\mathcal{N}_{z>1}>120$). Further increases in the sample size garner no appreciable improvement.

With $\approx 56$ \SNeIa\, the projected total error (statistical plus systematic) on $w_0$ and $w_a$ at the end of this supposed 10-year HST program would be approximately $\sigma_{w_{0}} < 0.18$ and $\sigma_{w_{a}} < 1.2$. These projected uncertainties from (mock) \SNeIa\ alone happen to be very
comparable to the total uncertainty that can presently be achieved by combining SN constraints with other cosmological probes. For example, \citep{SNLS33} use the SNLS3 SN compilation combined with measurements of the Hubble constant, the cosmic microwave background, and baryon acoustic oscillations. They find very similar uncertainties, $\sigma_{w_0} \sim 0.19$ and $\sigma_{w_a} \sim 1.1$.

\subsection{\SNIa\ evolution}

In the next decade, ground-based \SNIa\ samples will continue to grow, with wide-field surveys such as Pan-STARRS\footnote{Pan-STARRS: the Panoramic Survey Telescope and Rapid Response System; http://www.ps1sc.org/index.shtml}, DES\footnote{DES : the Dark Energy Survey; http://www.darkenergysurvey.org/} and LSST\footnote{LSST : the Large Synoptic Survey Telescope;http://www.lsst.org/lsst/} providing several thousand \SNIa\ at $z<1.5$. In this environment the real improvement in cosmological constraints from high-z \SNIa\ samples like CLASH+CANDELS may come from the ability to test for as-yet-unseen systematic biases in the \SNIa\ sample.

At low redshift $(z<1.5)$ the average \SNIa\ that we observe will have begun its life as a progenitor star with a main sequence (MS) mass $M_{MS}\sim 1 M_{\odot}$ or so.  However, at high redshift, when the universe was young, these low mass stars are still on the main sequence, so they cannot contribute to the observed high-z \SNIa\ population.  Thus the mean initial progenitor mass for the observable \SNIa\ population increases with redshift.   Following \citet{Riess06}, we adopt the progenitor mass - luminosity relationship of \citet{snevol} (hereafter $DHS2001$) as a plausible example of how this demographic shift could systematically bias the observed \SNIa\ distances and therefore cosmological parameters.  The $DHS2001$ model suggests that more massive \SNIa\ will be slightly fainter, which we might detect by seeing the average \SNIa\ peak magnitude shift systematically fainter than the baseline cosmological model as redshift increases. Using our mock high-z \SNIa\ samples, we can examine how such models for \SNIa\ evolution could be confronted by current and future surveys.

In addition to the MS lifetime, each \SNIa\ progenitor system must undergo some period of post-MS mass transfer. We consider two simplistic assumptions for this post-MS timescale: in case $1$ we assume that all \SNIa\ progenitors require $\tau=2.5$ Gyr after leaving the main sequence to reach the point of explosion;  for case $2$ we use $\tau=0.4$ Gyr. In reality there is a distribution of delay times, but these two extrema can serve to bracket a broad range of plausible progenitor models. To define the mean initial mass of a \SNIa\ progenitor as a function of redshift, we adopt a Salpeter initial mass function (IMF), and truncate the low-mass end by removing any stars with $M_{MS}<0.6 M_{\odot}$ (these would never accrete enough mass to reach the Chandrasekhar mass limit and explode) and also removing any for which the combined MS and post-MS timescales exceed the age of the universe since $z=11$. This is a slightly conservative prescription, as it implicitly assumes that all \SNIa\ progenitor stars  formed at $z=11$, although in fact many would have formed later and would therefore also still be on the MS or accreting towards the Chandrasekhar limit. We also truncate the high mass end of the IMF, removing stars with $M_{MS}>8.0 M_{\odot}$ (these would result in Core Collapse Supernovae instead of \SNIa).

Now with the doubly truncated IMF at each redshift $z$, we can compute the mean \SNIa\ progenitor mass at each redshift. The $DHS2001$ model predicts that an increase in the initial progenitor mass of $1 M_{\odot}$ would reduce the peak $B$ or $V$ band brightness by $\sim 0.03$ magnitudes $-$ without  a corresponding change in the color or light curve shape:  $dm_{B}/dM_{MS}\sim 0.03$. This converts the expected change in the mean initial mass into a systematic shift of the average peak magnitude for the high-z \SNIa\ population.

\begin{figure*}[htbp]
\centering
\includegraphics[width=7.7cm]{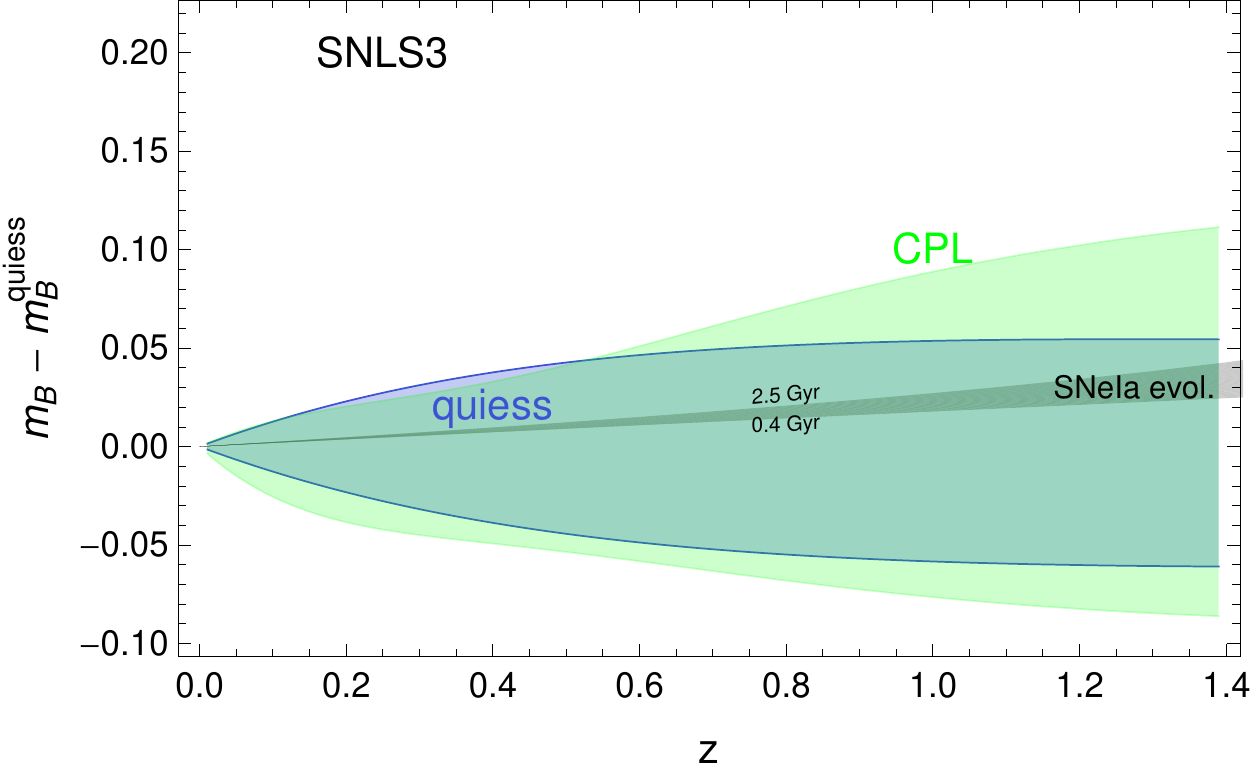}~~~
\includegraphics[width=7.7cm]{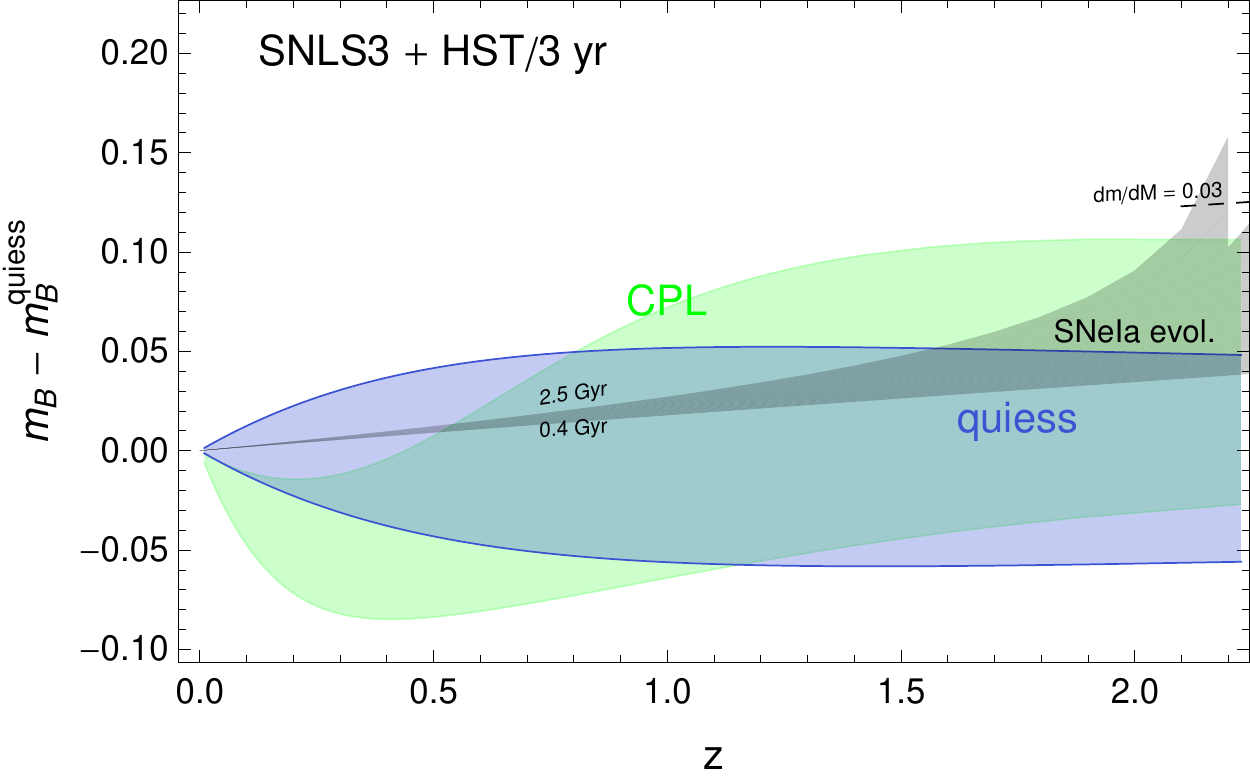}
~~~\\
\includegraphics[width=7.7cm]{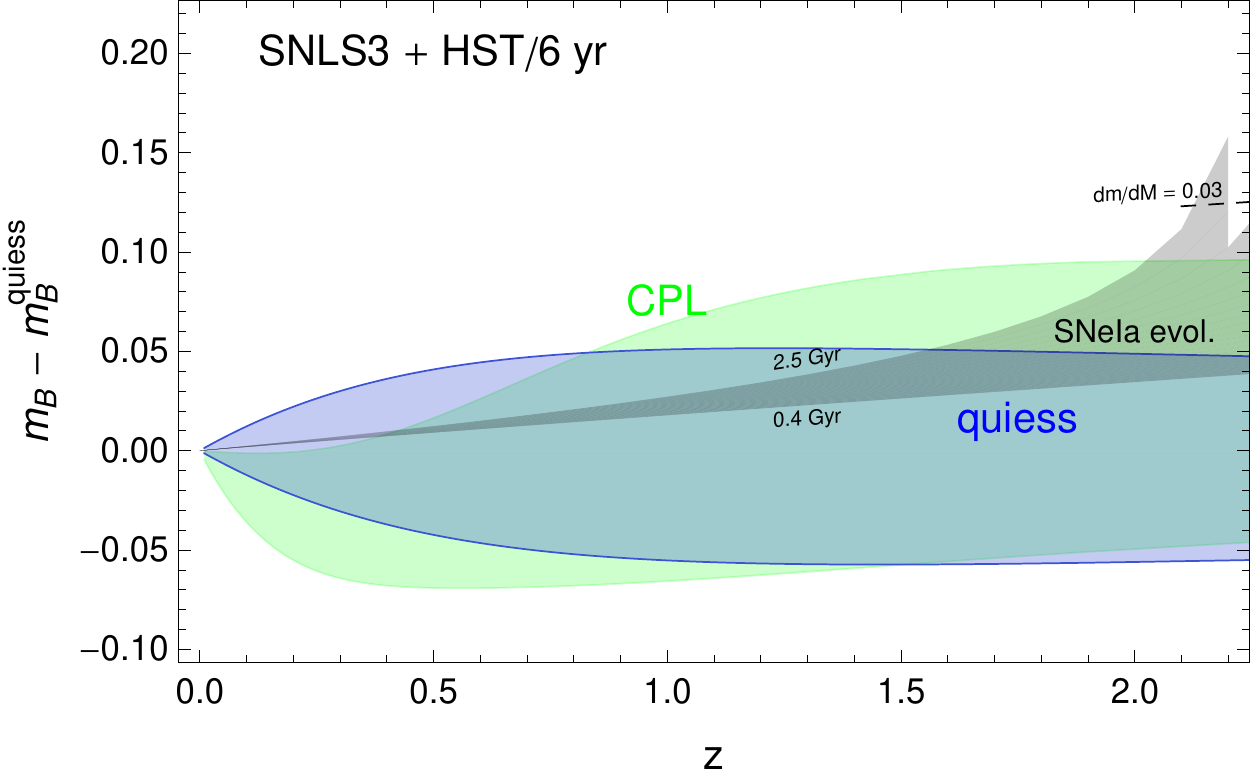}~~~
\includegraphics[width=7.7cm]{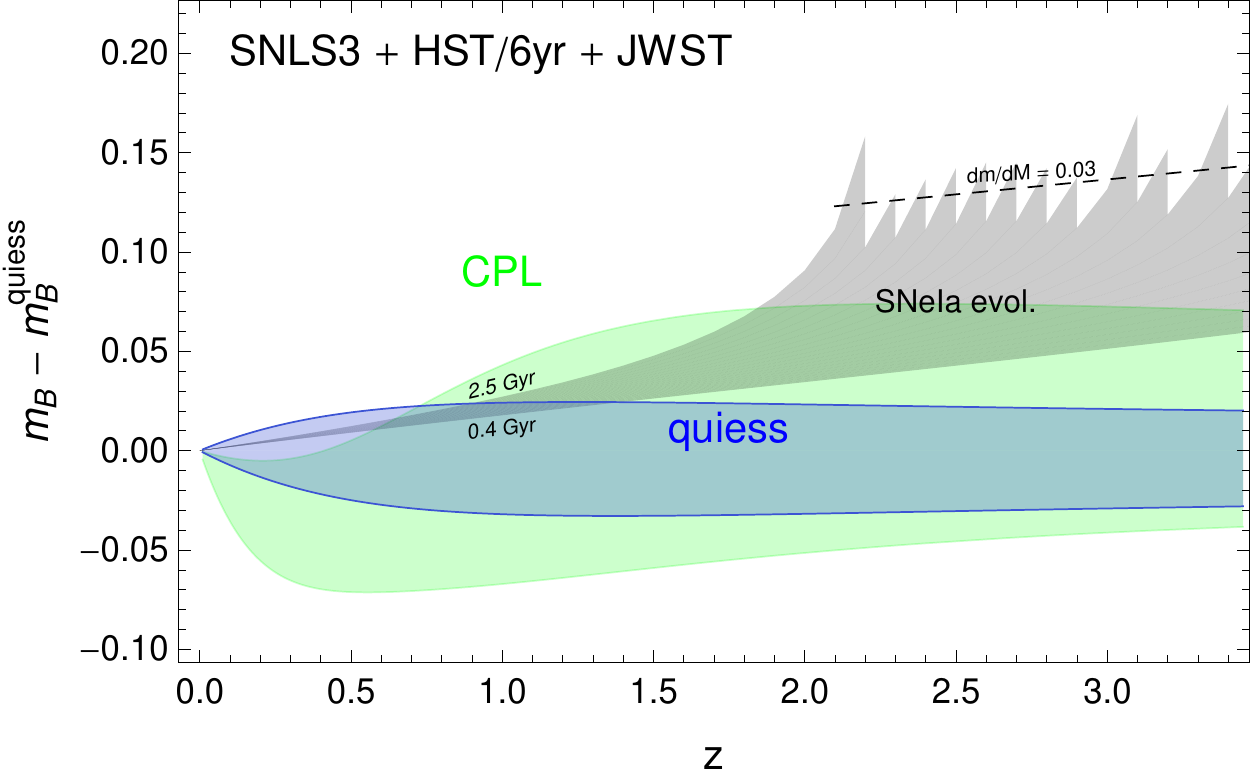}
\caption{Future tests for \SNIa\ evolution. Blue contours show the $1\sigma$ confidence levels derived from errors on the equation of state parameter $(w_{0})$ of quiessence models. Light green regions are the $1\sigma$ confidence levels derived from errors on $w_{0}$ and $w_{a}$ of CPL models. Grey regions denote a plausible range for \SNIa\ evolution effects, following \cite{snevol} and \cite{Riess06} and assuming $dm_{B}/dM_{MS}\sim 0.03$.}\label{fig:snevol}
\end{figure*}

To see if this effect would be distinguishable from dark energy evolution, we have analyzed our data sets with a quiessence cosmological model, which is cosmologically-time independent (constant equation of state). If we then calculate the residuals between \SNIa\ magnitudes and this reference cosmological model and, if there is any evolutionary effect, it should be detectable as an excess in magnitude. Nevertheless, we have to consider that we would have at least two possible evolutionary phenomena of different nature: one would be related to the \textit{cosmological} evolution (it would be, for example, the difference between a quiessence and a CPL model); and the other would be linked to the \SNIa\ evolution which we would like to detect. In Fig.~\ref{fig:snevol} we show blue contours representing the $1\sigma$ confidence levels derived from errors on the equation of state parameter ($w_{0}-$quiessence). Green contours show the difference between the CPL and quiessence models, representing the range of \SNIa\ magnitudes that is consistent with our current ignorance of dark energy evolution. Finally, the grey regions show the posited \SNIa\ evolutionary effect; if the grey region overlap the green CPL contours, then the two effects would be indistinguishable.

With present data, \SNIa\ evolution is clearly indistinguishable from a quiessence model if $dm_{B}/dM_{MS}\sim 0.03$. However, given the uncertainty regarding cosmological evolution, it is not possible to disentangle the two phenomena. The situation will not change when moving to a \minnum\ -like survey: in this case, even if the uncertainty from cosmological evolution is smaller, it is again too large to disentangle the two phenomena, and only \SNIa\ with $z>2.1$ could be helpful in this direction, even if they would not be enough numerous to be considered statistically significant. Adding more observations (\maxnum\ -like survey) improves the constraint on CPL parameters, thus lowering again the minimum redshift useful for the detection of \SNIa\ evolution ($z\gtrsim 2$).

In the bottom panel of Fig.~\ref{fig:snevol} we also plot a very likely future scenario. We can consider the impact of adding to the HST legacy with a high-z \SNIa\ survey using the \textit{James Webb Space Telescope}\footnote{http://www.jwst.nasa.gov/} (JWST); using the CLASH+CANDELS survey as a model, we estimate that a 3-year survey with JWST could yield $\approx 36$ \SNIa\ at $1.5<z<3.5$.  Adding this to the \maxnum\ sample further improves constraints on the CPL model (see last line in Table~\ref{tab:bestfit3}). The result is that \SNIa\ down to $z \approx 1.7$ could become useful probes for \SNIa\ evolution models, and even a model with $dm_{B}/dM_{MS}\sim 0.03$ could be, in principle, detectable. In closing, we note that these are likely conservative estimates: low-z \SNIa\ surveys (e.g. PS1, DES, LSST) should also improve the constraints on CPL models in the intervening decade, thus making the HST and JWST high-z \SNIa\ even more powerful as tests of \SNIa\ evolution models.

\section*{Acknowledgements}
We thank the anonymous referee for comments that greatly improved the quality and clarity of this work. Vincenzo Salzano, Ruth Lazkoz and Irene Sendra are supported by the Ministry of Economy and Competitiveness through research projects FIS2010-15492 and Consolider EPI CSD2010-00064 and by the Basque Government through research project GIU06/37. Support for Steven Rodney was provided by NASA through Hubble Fellowship grant $\#$HF-51312.01 awarded by the Space Telescope Science Institute, which is operated by the Association of Universities for Research in Astronomy, Inc., for NASA, under contract NAS 5-26555.

\end{document}